# Long-range p-d exchange interaction in a ferromagnet-semiconductor hybrid structure

V.L. Korenev[1,2,*], M. Salewski[2], I.A. Akimov[1,2,*], V.F. Sapega[1,3], L. Langer[2], I.V. Kalitukha[1], J. Debus[2], R.I. Dzhioev[1], D.R. Yakovlev[1,2], D. Müller[2], C. Schröder[4], H. Hövel[4], G. Karczewski[5], M. Wiater[5], T. Wojtowicz[5], Yu.G. Kusrayev[1], and M. Bayer[1,2]

[1]Ioffe Physical-Technical Institute, Russian Academy of Sciences, 194021 St. Petersburg, Russia

[2]Experimentelle Physik 2, Technische Universität Dortmund, D-44227 Dortmund, Germany

[3]Physical Faculty of St. Petersburg State University, 198504 St. Petersburg, Russia

[4]Experimentelle Physik 1, Technische Universität Dortmund, D-44227 Dortmund, Germany

[5]Institute of Physics, Polish Academy of Sciences, PL-02668 Warsaw, Poland

*Hybrid structures synthesized from different materials have attracted considerable attention because they may allow not only combination of the functionalities of the individual constituents but also mutual control of their properties. To obtain such a control an interaction between the components needs to be established. For coupling the magnetic properties, an exchange interaction has to be implemented which typically depends on wave function overlap and is therefore short-ranged, so that it may be compromised across the interface. Here we study a hybrid structure consisting of a ferromagnetic Co-layer and a semiconducting CdTe quantum well, separated by a thin (Cd,Mg)Te barrier. In contrast to the expected p-d exchange that decreases exponentially with the wave function overlap of quantum well holes and magnetic Co atoms, we find a long-ranged, robust coupling that does not vary with barrier width up to more than 10 nm. We suggest that the resulting spin polarization of the holes is induced by an effective p-d exchange that is mediated by elliptically polarized phonons.*

PACS: 85.75.–d, 78.67.–n

Exchange interactions are the origin for correlated magnetism in condensed matter with multiply-faceted behavior such as ferro-, antiferro- or ferrimagnetism. In magnetic semiconductors (SCs), the exchange occurs between free charge carriers and localized magnetic atoms [1, 2, 3, 4] and is determined by their wavefunction overlap. To assess and control this overlap, hybrid structures consisting of a ferromagnetic (FM) layer and a semiconductor quantum well (QW) are appealing objects because they allow wavefunction engineering. Furthermore, the mobility of QW carriers may not be reduced by inclusion of magnetic ions in the same spatial region.

More specifically, for a two-dimensional hole gas (2DHG, the *p*-system) in a QW the overlap of the hole wavefunction with the magnetic atoms in a nearby ferromagnetic layer (the *d*-system) is believed to result in a *p-d* exchange interaction [5, 6, 7, 8]. This exchange interaction may cause strong coupling between the SC and FM spin systems [9], through which the ferromagnetism of the unified system, as evidenced by its hysteresis loop, for example, can be tuned. In particular, the 2DHG spin system becomes polarized in the effective magnetic field of the *p-d* exchange [5, 8]. Recently [10], it was shown that additionally to this equilibrium 2DHG polarization there is an alternative mechanism involving spin-dependent capture of charge carriers from the SC into the FM. For ferromagnetic (Ga,Mn)As on top of an (In,Ga)As QW, electron capture induces their spin polarization in the QW, representing a *dynamical* effect in contrast to the exchange-induced equilibrium polarization.

Here we study a different FM/QW hybrid consisting of a Co-layer and a CdTe-QW, i.e. a II-VI semiconductor, separated by a nanometer-thick barrier. Due to the negligible hole tunneling, this hybrid combination shows mostly the quasi-equilibrium ferromagnetic proximity effect due to the *p–d* exchange interaction between Co atoms and CdTe-heavy holes. Surprisingly, however, the proximity effect, measured through the spin polarization of the heavy holes in the QW, is almost constant over large distances up to 14 nm spacer width. In contrast, for conventional *p-d* exchange via wavefunction overlap an exponential decay with barrier width with a characteristic decay length of about a nanometer would be expected. This novel exchange coupling effect is therefore truly long-range, which is highly advantageous because it is robust with respect to hybrid interface variations. As possible origin of this long-range proximity effect we suggest an effective *p-d* exchange interaction mediated by elliptically polarized phonons emitted from the FM into the QW.

**Ferromagnet-induced circular polarization of the QW photoluminescence**

We study a set of gradient structures composed of a Co-layer, a (Cd,Mg)Te-spacer (cap), and a CdTe-QW followed by another (Cd,Mg)Te-barrier that was grown on top of a (100)-oriented GaAs substrate (see Fig.1a for a general design of the structures). In the structure 1 the (Cd,Mg)Te spacer was grown with the main shutter moving in front of the substrate in order to have its thickness $d_s$ increasing continuously from 5 nm to 15 nm over a lateral structure dimension of 5 cm (wedge shape). On top of this semiconducting sequence a 7 nm thick Co film was deposited. The semiconducting parts of the other two structures are similar to those of structure 1 except for the spacer thickness $d_s$, which now was varied from zero to 10 nm and from zero to 14 nm for structure 2 and 3, respectively, over 3 cm lateral dimension, followed by an area with the constant thickness $d_s = 50$ nm. Furthermore, for these samples, the Co film thickness $d_{Co}$ varied in discrete steps of 1 nm from zero to 6 nm along the direction perpendicular to spacer thickness gradient (Fig. 1a), followed by an area with the Co-film having constant thickness of 16 nm. The Co layers are semitransparent for the exciting laser light. Atomic force microscopy shows a non-uniform Co growth with islands having tens of nanometers lateral sizes and height of about 5 nm (Fig. 1b). In the following we present the main experimental data for structure 2 if not mentioned otherwise. We note that the main observations concerning the proximity effect are similar for all structures.

The samples are illuminated with continuous or pulsed laser light, being either linearly (π) polarized or unpolarized through a depolarizing wedge. An out-of-plane magnetization component can be induced by applying an external magnetic field in Faraday geometry $\mathbf{B}_F$ (longitudinal field parallel to the structure growth axis $\mathbf{z}\|[001]$). The band structure of the hybrid system is sketched in Fig. 1c. For photon energies smaller than the band gap energy $E_g^{CdMgTe} = 2$ eV of the barrier, only the QW is excited (Fig. 1a). The photoluminescence (PL) emitted by the QW is analyzed with respect to its circular polarization $\sigma^+$ and $\sigma^-$, to determine the circular polarization degree $\rho_c^\pi(B_F)$ of the PL under π-excitation.

Figure 1d shows half the difference $\delta\rho_c^\pi$ between the polarization degrees recorded for $B_F = +10$ and $-10$ mT (blue dots) at $T = 2$ K for a spacer thickness of $d_s = 10$ nm in comparison with a PL spectrum (black line). Further on we will often use the polarization difference $\delta\rho_c^\pi(B_F) = [\rho_c^\pi(+B_F) - \rho_c^\pi(-B_F)]/2$ as a measure of the field-induced signal. Two main lines centered at 1.63 eV (X-line) and ~1.60 eV (e–$A^0$-line) with a wide flank towards lower energies contribute to the PL. We assign the X-line to the heavy-hole exciton transition of the QW. The e–$A^0$ line, on the other hand, is associated with electrons recombining with holes bound to acceptors. This assignment is supported by time-resolved PL data, where the X-line decays within 50 ps while the e–$A^0$ PL is observed during much longer times of nanoseconds (Fig. 2a). From the typical exciton binding energy of 10 meV and the 20-30 meV distance between the two lines one deduces a binding energy of $E_{A^0} \approx 30\text{-}40$ meV, which agrees well with the values for a hole bound to a shallow acceptor in CdTe [11]. The degrees of circular PL polarization $\rho_c^\pi(B_F)$ vary considerably for $B_F = +10$ and $-10$ mT (Fig. 1d) with a maximum difference at the e–$A^0$ line. A photoluminescence excitation spectrum reveals no strong dependence of $\rho_c^\pi(B_F)$ on the laser photon energy in the range of photon energies from 1.62 - 1.69 eV. Figure 1e shows the magnetic field dependence of $\rho_c^\pi(B_F)$ for excitation at $h\nu_{exc} = 1.69$ eV and detection at $h\nu_{PL} = 1.59$ eV. Up-and-down magnetic field scans show a non-hysteretic dependence of $\rho_c^\pi(B_F)$ saturating at a field strength of $|B_{sat}| = 20$ mT with a polarization amplitude $A = \delta\rho_c^\pi(B_{sat}) \approx 4\%$, being an order of magnitude larger than without Co-layer (see Supplementary Figure S3c). Fig. 1e also shows the dependence $\rho_c^\pi(B_F)$ detected at $h\nu_{PL} = 1.494$ eV, corresponding to the e–$A^0$ PL transition from the GaAs substrate (the red triangles). In contrast to the QW emission, we observe a much smaller polarization degree without any saturation. Therefore the PL polarization from the CdTe QW is clearly induced by the FM. We stress that magnetic circular dichroism in the FM layer and diffusion of magnetic atoms into the QW region cannot cause the spin polarization (see Supplementary Section A; the origin of the ferromagnet responsible for the polarization of the QW PL is discussed in Supplementary Section B).

Time-resolved PL (TRPL) gives insight into the kinetics of the spin polarization after linearly polarized excitation. Spectrally resolved intensity transients are shown in Fig. 2a. Figure 2b shows the temporal dependence of the FM-induced half the difference $\delta\rho_c^\pi(t)$ between circular PL polarizations in magnetic fields of +40 mT and −40 mT, applied in Faraday geometry: $\delta\rho_c^\pi(t)$ continuously increases with time: initially being non-polarized, the photo-excited carriers acquire spin polarization with a characteristic rise time of $\tau_{fm} \approx 2$ ns, obtained from exponential fit (dashed line). In combination with the decay times of 50 ps for the X line and a few nanoseconds for the e–$A^0$ line this allows one to explain the absence of polarization for the X-line: the exciton does not live long enough to acquire polarization from the FM (another possibility will be discussed below).

To understand the origin of the FM-induced PL polarization from the QW we have to address the following questions: (i) Are the QW electrons or holes polarized? (ii) Does the spin-dependent capture of carriers from the QW into the FM play a role? (iii) Is there an effective exchange magnetic field from the FM polarizing the QW carriers?

### (i) Ferromagnet-induced spin polarization of the QW heavy holes

As discussed, application of a longitudinal magnetic field $B_F$ causes a polarization of the PL. Now we apply an additional magnetic field $B_V$ in Voigt geometry with the goal to depolarize the FM-induced PL polarization (see scheme in Fig. 3a). Figure 3b shows that for $|B_F| = 4$ mT the half polarization difference $\delta\rho_c^\pi$ of the QW PL is about 1% and is *not sensitive* to the magnetic field $B_V$ up to 20 mT. This means that the out-of-plane $M_z$ component of the FM remains fixed in this range of Voigt-fields. However, this is not the only consequence. The data also indicate heavy hole polarization as source of the non-zero $\delta\rho_c^\pi$. Indeed, if $\delta\rho_c^\pi$ would arise from electron spin orientation, the Hanle effect [12] should be observed for the electron spins. Electrons spins are polarized along the z-axis, and the magnetic field $B_V$ induces Larmor precession about $B_V$. The frequency of precession should decrease the z-component of the electron spin, leading to a PL depolarization $\rho_c^\pi(B_V)$ in complete analogy with the Hanle effect under optical injection of the electron spin. Indeed, under circularly polarized excitation the degree of polarization $\rho_c^\sigma(B_V)$ decreases with $B_V$ (green solid circles in Fig. 3c) with a halfwidth $B_{1/2} \approx 15$ mT. However, such a dependence is absent for $\rho_c^\pi(B_V)$ in the same field range (Fig. 3b). Moreover, Fig. 3c also shows (the magenta triangles) that application of 4 mT longitudinal field does not change the Hanle curve. It means that the magnetization of the FM does not create a noticeable magnetic field (e.g., due to *s-d* exchange) which would influence the Larmor precession of the electron spin. Therefore, the out-of-plane magnetization $M_z$ of the FM does neither orient electron spins, nor does it affect their Larmor precession frequency. We conclude that the FM-induced polarization $\rho_c^\pi$ is directly related to the spin polarization of heavy holes: the in-plane g-factor of heavy holes is close to zero, so that the $B_V$ field cannot depolarize them.

TRPL data allow one to determine the electron and hole spin dynamics for comparison with the ~1 ns kinetics of the proximity effect. The bottom panel of Fig. 2c shows the optical orientation signal $\bar{\rho}_c^\sigma(t) = [\rho_c^\sigma(+B_F,t) + \rho_c^\sigma(-B_F,t)]/2$ at $|B_F|$=40 mT, averaged over the magnetic field direction, to exclude any contribution of the FM-induced polarization. One finds two decay components with strongly different spin relaxation times $\tau_{se}$ and $\tau_{sh}$. The shorter one $\tau_{sh} = 0.12$ ns corresponds to holes, whereas the much longer time $\tau_{se} > 20$ ns corresponds to electrons [10]. It is important to note that the zero-field decay of $\bar{\rho}_c^\sigma(t)$ in Fig. 2c contains an additional contribution $\tau_{deph} \approx 0.3$ ns due to electron spin dephasing in the randomly oriented stray fields ~15 mT of the FM. The stray fields also determine the 15 mT width of the Hanle curve (Supplementary Section A2). This dephasing affects the spin dynamics and leads to the faster decay of $\bar{\rho}_c^\sigma(t)$ at $B_F = 0$. Application of a $B_F = 40$ mT Faraday-field suppresses the stray field impact. The comparison of $\bar{\rho}_c^\sigma(t, B_F = 40\,\text{mT})$ with the kinetics $\delta\rho_c^\pi(t)$ shows that the FM-induced polarization kinetics is much faster ($\tau_{fm} \approx 2$ ns) than the electron spin kinetics $\tau_{se} > 20$ ns in agreement with the hole spin flip being much faster than that of the electron [12].

However, it is considerably slower than the hole spin-flip time of $\tau_{sh} = 0.15$ ns. This faster spin relaxation of the optically orientated holes can be explained by the higher laser photon energy (1.70 eV) by which free holes are excited. The optically excited hot holes become depolarized before they are trapped by acceptors.

Here we want to recall the unusual proximity effect shown in Fig. 1d: only the e-A$^0$ transition shows a FM-induced polarization. Hence $\rho_c^\pi(t)$ reflects the spin kinetics of the holes bound to acceptors, whose spin-flip time can be considerably longer than $\tau_{sh}$. Further evidence (Fig. 3d, upper panel) of the heavy hole polarization comes from the non-oscillating signal $\delta\rho_c^\pi(t, B_F = 10 \text{ mT})$ increasing with time in spite of the large magnetic field $B_V = 96$ mT applied in Voigt geometry. In contrast, the optical orientation signal $\bar{\rho}_c^\sigma(t, B_F = 10\,mT)$ oscillates in time according to the electron g-factor |g$_e$|=1.2 and goes to zero (Fig. 3d, bottom panel).

We conclude that: 1) the out-of-plane magnetization component of the FM is robust in magnetic fields $B_V$ <100 mT, and 2) the FM-induced PL polarization $\rho_c^\pi(B_F)$ in longitudinal field originates from the spin polarization of heavy-holes bound to acceptors.

**(ii) Spin-dependent capture as possible source of non-equilibrium QW spin polarization**
According to [10] the observed proximity effect may be due to spin-dependent carrier capture from the semiconductor into the FM. In this case, the total PL intensity $I^{\sigma\pm}$ (the sum of the right- and left-handed circular PL components) in longitudinal magnetic field would depend on the helicity $\sigma^\pm$ of the exciting light. The intensity modulation parameter $\eta = (I^{\sigma+} - I^{\sigma-})/(I^{\sigma+} + I^{\sigma-})$ is essentially determined by the selection rules for the involved optical transitions $P_i(h\nu)$ at the laser energy $h\nu$. According to [10], $\eta(B_F) = P_i \cdot \rho_c^\pi(B_F)$ is equal to $\rho_c^\pi(B_F)$ for heavy hole exciton excitation when $P_i \approx 1$. Figure 4b shows the modulation parameter $\eta(B_F)$ when the laser photon energy was tuned to the X resonance at 1.62 eV and detected at the e-$A^0$ transition (1.60 eV). $\eta(B_F)$ is several times smaller than the polarization degree $\rho_c^\pi(B_F)$. In fact, $\eta(B_F)$ is weak over the whole range of laser energies 1.62-1.70 eV. We conclude that spin-dependent capture by the FM is not the main source of the FM induced hole spin polarization.

**(iii) Role of p-d exchange in spin polarization of QW heavy holes**

Heavy hole spin polarization in the effective magnetic field of the *p-d* exchange interaction with the magnetic atoms was predicted in Ref. [5]. The equilibrium polarization of the holes, when the FM magnetization is saturated ($M = M_s$), obeys the Curie law for non-degenerate statistics

$$P_s(T) = \frac{N_{+3/2} - N_{-3/2}}{N_{+3/2} + N_{-3/2}} \approx \frac{\Delta E_{ex}(M_s)}{2k_B T} \tag{1}$$

Here $k_B$ is the Boltzmann constant, $\Delta E_{ex}(M_s)$ is the spin splitting of the heavy holes in the FM exchange field, $N_{+3/2}(N_{-3/2})$ is the concentration of QW heavy holes with momentum projection $+3/2(-3/2)$ onto the growth direction and we assume that $P_s(T) \ll 1$. One can expect from Eq.(1) that the hole spin polarization decreases with increasing $T$. The experiment confirms this tendency (Fig. 4a). An estimation for $T = 10$ K shows that a few percent spin polarization corresponds to the splitting $\Delta E_{ex}$ of about 50 µeV.

The *p-d* exchange from the overlap of the QW holes and Co *d*-shells is expected to scale like $\Delta E_{ex}(d_s) \sim \exp(-d_s/d_0)$ due to wavefunction penetration through the rectangular potential barrier. The same exponential behavior holds for the tunneling rate with the characteristic barrier width $d_0 = \hbar/2\sqrt{2m_h \Delta}$ which is less than 1 nm for the heavy holes mass $m_h > 0.1 m_0$ and the barrier height $\Delta = 0.1$ eV. Thus the two possible coupling effects discussed above are definitely expected to be short-range. At first sight, this is in agreement with the steep decrease of PL intensity due to carrier tunneling from the QW into the Co with decreasing barrier thickness (red squares Fig. 1f), from which we find $d_0$ =1.6 nm. However, the spin polarization does not follow a similar steep decrease at all. The green circles in Fig. 1f show that $\rho_c^\pi(d_s)$ at $B_F$ = 40 mT is roughly constant and almost independent of $d_s$. Furthermore and maybe even more surprising, the proximity effect observed in our case is *long-ranged* and persists over more than 14 nm, in contrast to our expectation from wavefunction overlap.

It should be noted that in other systems than the one studied here a long-range proximity effect may occur, e.g. due to the spin-dependent capture by the FM. Previously this effect was observed in a ferromagnetic GaMnAs/QW hybrid for electrons, and not for holes [10]. The individual capture rates $\gamma_+(\gamma_-)$ of electrons with spin up (down) have also an exponential tunneling rate through the nonmagnetic spacer. The non-equilibrium polarization in the QW, however, is determined by the ratio $\gamma_+/\gamma_-$, so that the exponential dependence is cancelled. Here the spin dependent capture is weak in absolute terms, as it is related to holes, and the proximity effect comes mainly from the *p-d* exchange interaction. Therefore we can conclude that our results indicate a novel – long-range – mechanism of *p-d* exchange coupling.

**Origin of long-range *p-d* exchange coupling**

In the following we discuss three possibilities for the observed long-range coupling: a) magnetization of holes by the stray fields of the FM film; b) resonant tunneling through deep centers in the CdMgTe spacer; c) spin polarization of holes by elliptically polarized phonons emitted from the FM towards the QW.

**a. Effect of FM stray magnetic fields on spin polarization of charge carriers**

It is known that stray magnetic fields created in a SC by a nearby FM [13, 14] can have large penetration depths and lead to dephasing (with characteristic time $\tau_{deph}$) of electron spin Larmor precession (Fig. 3d). The stray field averaged over a region much larger than the size of a magnetic domain is $4\pi M_s d_{Co}/L \sim$ 0.02 mT for a Co film thickness $d_{Co} = 5$ nm and a sample size of $L \approx 5$ mm. The hole spin polarization in a field of this strength is negligible. Larger local stray fields $\sim 4\pi M_s d_{Co}/w$ appear on length scales comparable with the domain size $w$. The stray field measured by optical orientation is about 20 mT (Supplementary Section A2) and cannot provide a notable spin polarization. Therefore stray fields cannot explain the observed hole spin polarization.

**b. Resonant tunneling through deep centers in the (Cd,Mg)Te spacer**

Another possible reason is related to resonant transfer of spin coupling through overlapping paramagnetic deep centers in the (Cd,Mg)Te barrier. Assuming a localization radius of the center of about $a_t \sim 1$ nm one can estimate the concentration required to make such a coupling efficient to be $N_t \sim 1/a_t^3 \sim 10^{21}$ cm$^{-3}$. Such a large concentration of paramagnetic centers was never reported for (Cd,Mg)Te. Moreover, its presence should significantly affect the Larmor precession frequency of the CdTe QW electrons, which was not observed. Therefore we consider this possibility as highly unlikely (see also Supplementary Section A3).

**c. Spin polarization of holes by elliptically polarized phonons**

The long-range *p-d* exchange between FM and SC may be mediated by elliptically polarized phonons generated in the ferromagnetic layer. Below we consider acoustic phonons as an example. Coupling with elliptically polarized optical phonon modes [15] is discussed in Supplementary Section C. Phonons propagating through a FM along the magnetization direction **m** = **M**/$M_s$ are elliptically polarized due to the strong hybridization of the acoustic phonon (linear in momentum) and spin wave (quadratic in momentum) modes near the crossing points of their dispersions, i.e. for a phonon-magnon resonance [16]. There are two crossing points at frequencies of $\omega_1 \sim 10^{10}$ s$^{-1}$ and $\omega_2 \sim 10^{12}$ s$^{-1}$ with two orthogonal phonon polarizations per crossing [16]. Only the phonon mode with polarization vector rotating in the same direction as the magnetization vector in the spin wave participates in the coupling, while the other one couples only weakly to magnons [17]. The elliptically polarized phonons transfer angular momentum through the FM.

We suggest that such elliptically polarized phonons can affect the hole spin system when transmitted through the interface between FM and SC with not too different acoustic impedances. In this case elliptically polarized acoustic modes will propagate and hit the QW influencing the confined hole spins, which resembles a proximity effect for phonon angular momentum. The effect may obviously occur over much larger spacer thicknesses than any wavefunction overlap-based mechanism because there is no energy barrier for phonons. Any hole spin-phonon coupling is expected to change the energy of the heavy hole spin subbands due to the strong spin-orbit interaction in the valence band. To the best of our knowledge this effective *p-d* exchange coupling has not been discussed previously in literature and may be considered as phonon analog of the ac Stark shift well known in optics [18], in which the action of circularly polarized electromagnetic wave in a medium is equivalent to an effective magnetic field.

Let us discuss the effect in a bit more detail by assuming an elliptically polarized transverse phonon that propagates along **M** (axis $\mathbf{z} \parallel [001]$) so that the angular momentum of the phonon is parallel to **M**. Furthermore, we assume for illustrational purposes that only the $\sigma_{phon}^{-}$ polarized phonon mode exists and propagates through the FM-SC interface (Fig. 5). The interaction of the hole spin with this rotary lattice oscillation couples the ground state $|+3/2, N\rangle$ of the $+3/2$ heavy hole in the quantum well in presence of $N$ phonons with the excited state $|+1/2, N-1\rangle$ of $+1/2$ light holes in the presence of $N-1$ phonons, conserving thereby angular momentum (see Supplementary Section C for details). This leads to the "phonon ac Stark effect" – a spin-dependent shift of hole spin levels. The shift will be the stronger the closer the phonon energy $\hbar\omega_k \approx \Delta_{\ell h}$ is to heavy-light hole spin splitting $\Delta_{\ell h}$ (energy level diagram in Fig. 5). The energy shift $E_{\pm 3/2}$ will be different for the $\pm 3/2$ levels. This dynamic, phonon-mediated interaction induces therefore an effective *p-d* exchange coupling between FM and QW, which can be written as

$$V_{eff}^{pd} = -\frac{1}{2} J_{eff} m_z \sigma_z \qquad (2)$$

with the Pauli matrix $\sigma_z$ in the basis of heavy-hole states $\pm 3/2$, and $J_{eff}$ being the effective *p-d* exchange constant. The calculation of the "phonon ac Stark shift" of the QW hole levels requires precise knowledge of many quantities - the interaction parameters, phonon propagation directions, etc. - and is beyond the scope of the present work. However, we expect that the hole spin polarization changes sign under reversal of $M_z$, and the splitting $\Delta E_{ex} = |E_{+3/2} - E_{-3/2}|$ becomes larger for heavy-light hole splittings $\Delta_{\ell h} \approx \hbar\omega_1, \hbar\omega_2 < 1$ meV. Correspondingly, the splitting $\Delta E_{ex}$ is expected to be larger for the acceptor bound holes $A^0$ with $\Delta_{\ell h}^A \sim 1$ meV [19], than for the free QW holes with $\Delta_{\ell h}^{free} \sim 10$ meV. This is another possible explanation for detection of the proximity effect at the e-$A^0$ transition and not at the X transition. The absence of a FM-induced spin splitting of QW electrons (and the related precession corresponding to the Larmor frequency) can be also explained by the small spin-orbit interaction in the conduction band.

All the discussed features are qualitatively consistent with the experimental data. Therefore we come to the following basic *conjecture*: *the long-range p-d exchange coupling is* *mediated by elliptically polarized phonons*. A direct observation of elliptically polarized phonons in semiconductors could have profound consequences for the physics of spin systems in general, and not only for hybrids. Firstly, it could enable one to control the spin levels through the phonon analog of the optical ac Stark effect [18]. Secondly, one could show phonon-assisted spin pumping: the absorption of circularly polarized phonons would create spin orientation of holes in the valence band, similar to the well-known interband optical pumping in solids [20] and gases [21].

**Acknowledgments:**

We acknowledge support by the Deutsche Forschungsgemeinschaft and Russian Foundation for Basic Research in the frame of the ICRC TRR 160, the Government of Russia via project N14.Z50.31.0021, and the Program of Russian Academy of Sciences. V.L.K. acknowledges the Deutsche Forschungsgemeinschaft financial support within the Gerhard Mercator professorship program. The research in Poland was partially supported by the National Science Center (Poland) under grants nos. DEC-2012/06/A/ST3/00247 and DEC-2014/ST3/266881. T. W. acknowledges also the support from the Foundation for Polish Science through the International Outgoing Scholarship 2014.

**Additional information**

The authors declare no competing financial interests.

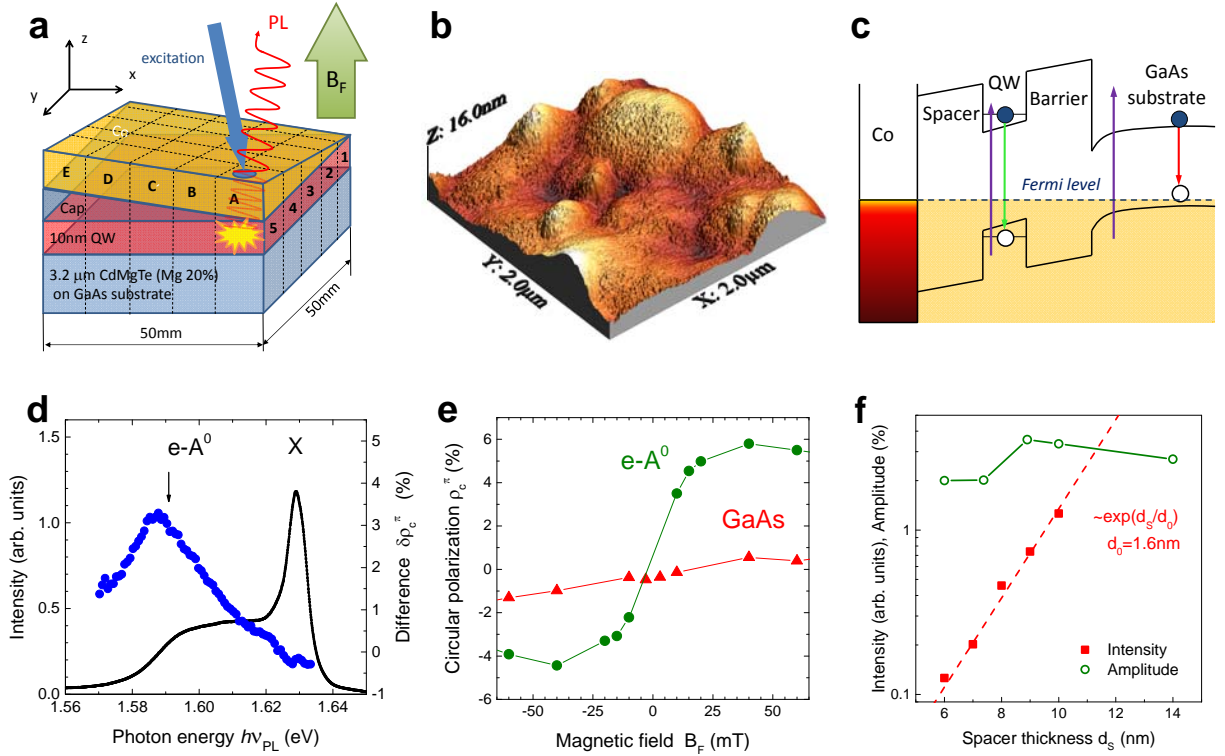

**Figure 1. Ferromagnetic-induced proximity effect.** (a) Scheme of optical excitation of double gradient structure 2. (b) AFM image of structure 2 with Co film thickness $d_{Co} = 5$ nm. (c) Band structure of investigated structures. Violet arrow indicates the excitation energy, while green and red arrows correspond to PL from QW and GaAs substrate, respectively. (d) PL spectrum (black line) and half the difference $\delta\rho_c^\pi = [\rho_c^\pi(+B_F) - \rho_c^\pi(-B_F)]/2$ (blue dots) between circular polarizations $\rho_c^\pi$ measured for $B_F = -10/+10$ mT at $T = 2$ K, linearly polarized pulsed excitation with photon energy 1.69 eV and excitation density 20 W/cm$^2$. The spacer thickness $d_s = 10$ nm and Co thickness $d_{Co} = 4$ nm; (e) The dependence $\rho_c^\pi(B_F)$ detected at the lower energy flank of e-A$^0$ transition from the QW around 1.59 eV as it is indicated with vertical arrow in panel (d) (green circles) and GaAs substrate at 1.49 eV (red triangles); up-down field scan revealed no hysteresis. (f) Dependence of exciton PL intensity on spacer thickness $d_s$ for $d_{Co} = 4$ nm (red squares). Dashed line is an exponential fit exp($-d_s/d_0$) with characteristic length $d_0 = 1.6$ nm. Open circles compare the spacer dependence of the proximity effect amplitude $A \equiv \rho_c^\pi(B_F = 40\text{ mT})$, which appears to be constant.

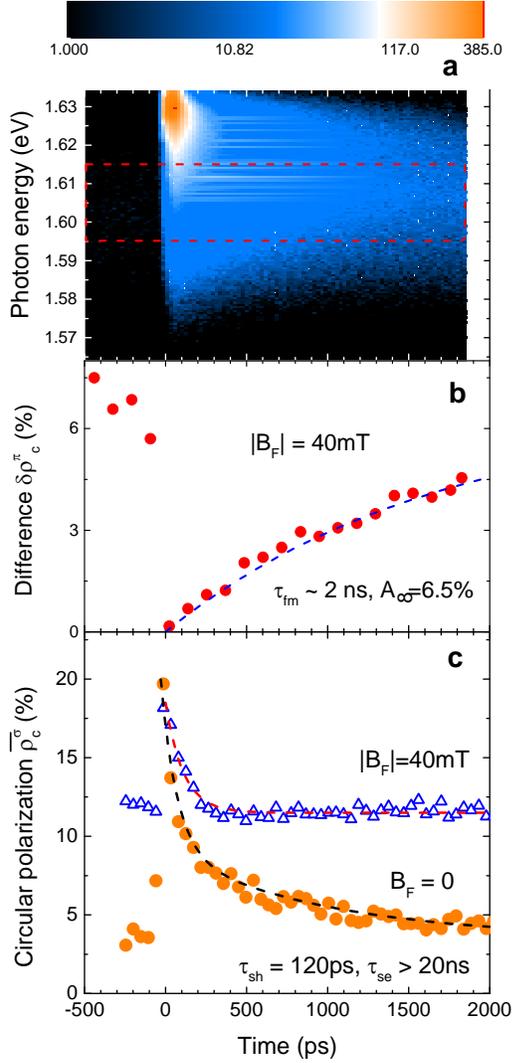

**Figure 2**: **Time-resolved spin dynamics.** (a) Contour plot of PL intensity decay from the QW. The frame indicates the region of interest for polarization measurements which corresponds to e–A$^0$ transition. (b) Dynamics of FM induced proximity effect. Time evolution of half the difference of circular polarization degree $\delta\rho_c^\pi = [\rho_c^\pi(+B_F) - \rho_c^\pi(-B_F)]/2$ for $|B_F| = 40$ mT under linear polarized excitation is shown. Dashed line is a fit with $\delta\rho_c^\pi(t) = A_\infty[1-\exp(-t/\tau_{fm})]$ with $\tau_{fm} \approx 2$ ns and $A_\infty = 6.5\%$; (c) Optical orientation kinetics. Circular polarization under excitation by $\sigma^+$-light $\bar{\rho}_c^\sigma(t) = [\rho_c^\sigma(+B_F,t) + \rho_c^\sigma(-B_F,t)]/2$ are shown for $|B_F| = 40$ mT and $B_F = 0$. Red dashed line results from double exponential fit of the data for $|B_F| = 40$ mT with $\bar{\rho}_c^\sigma(t) = \rho_{0,h}\exp(-t/\tau_{sh}) + \rho_{0,e}\exp(-t/\tau_{se})$, where $\rho_{0,h} = 7\%$, $\tau_{sh} = 0.12$ ns, $\rho_{0,e} = 11.5\%$ at $|B_F| = 40$ mT and $\tau_{se} > 20$ ns.

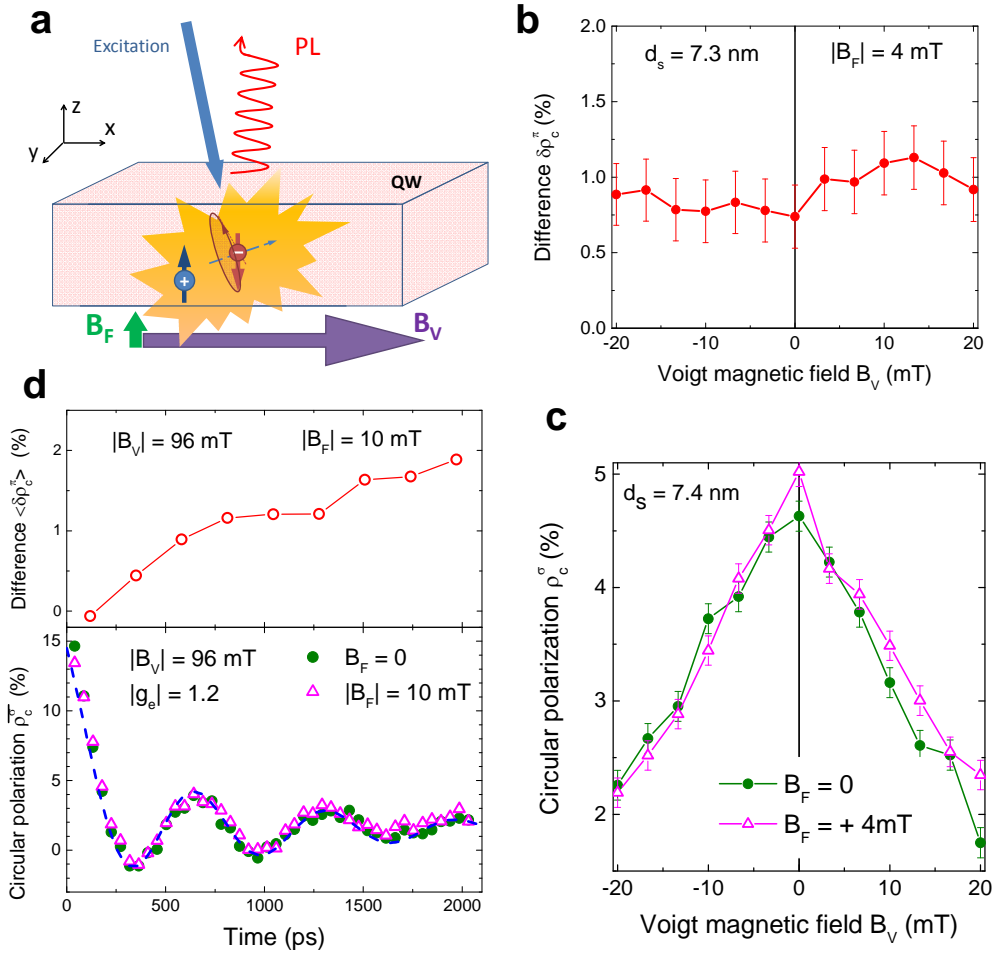

**Figure 3. Ferromagnetic-induced spin polarization of the QW heavy holes bound to acceptors.** (a) Sketch of the experiment in crossed magnetic fields. (b) Sample 1. Dependence $\delta\rho_c^\pi(B_V)$ in magnetic field in Voigt geometry in the presence of a longitudinal field $|B_F| = 4$ mT. (c) Hanle effect $\rho_c^\sigma(B_V)$ in the absence of the longitudinal field (green circles) and in the presence of $B_F = +4$ mT (magenta triangles). (d) Sample 2, $d_s = 10$ nm. Upper panel: Kinetics of $\langle\delta\rho_c^\pi(t)\rangle = [\delta\rho_c^\pi(t,+B_V) + \delta\rho_c^\pi(t,-B_V)]/2$ of proximity effect at $|B_V| = 96$ mT, $|B_F| = 10$ mT. Bottom panel: Kinetics $\overline{\rho}_c^\sigma(t)$ of optical orientation in magnetic field $|B_V| = 96$ mT, $B_F = 0$ (green circles), $|B_F| = 10$ mT (magenta triangles). Dashed line is the fit with $\overline{\rho}_c^\sigma(t) = \rho_{0,h}\exp(-t/\tau_{sh}) + \rho_{0,e}\exp(-t/\tau_{deph})\cos(g_e\mu_B B_V t/\hbar) + C$ with $\rho_{0,h} = 8\%$, $\tau_{sh} = 0.15$ ns, $\rho_{0,e} = 5\%$, $|g_e| = 1.2$, $\tau_{deph} = 1$ ns, $C = 1.5\%$. $T = 2$ K.

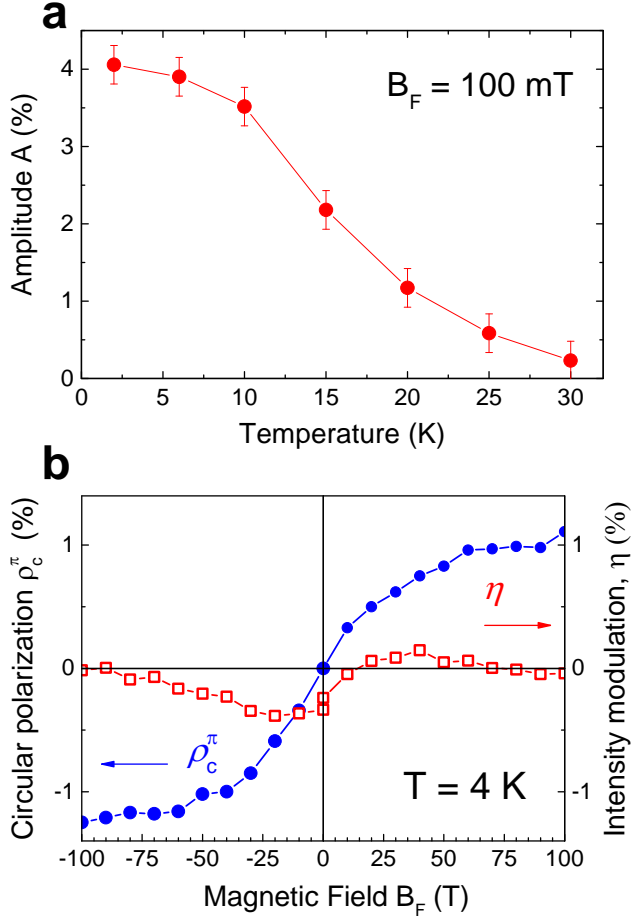

**Figure 4. Effective *p-d* exchange interaction between FM and QW heavy holes.** (a) Temperature dependence of the saturation amplitude in sample 1. (b) Sample 1. Resonant excitation of exciton (1.619 eV), detection e-A$^0$ (1.602 eV). The dependence of intensity modulation parameter $\eta(B_F)$ (red squares, right axis, laser with alternate helicity without polarizers in the detection channel) is compared with FM-induced polarization $\rho_c^\pi(B_F)$ (blue circles, left axis, $\pi$- excitation).

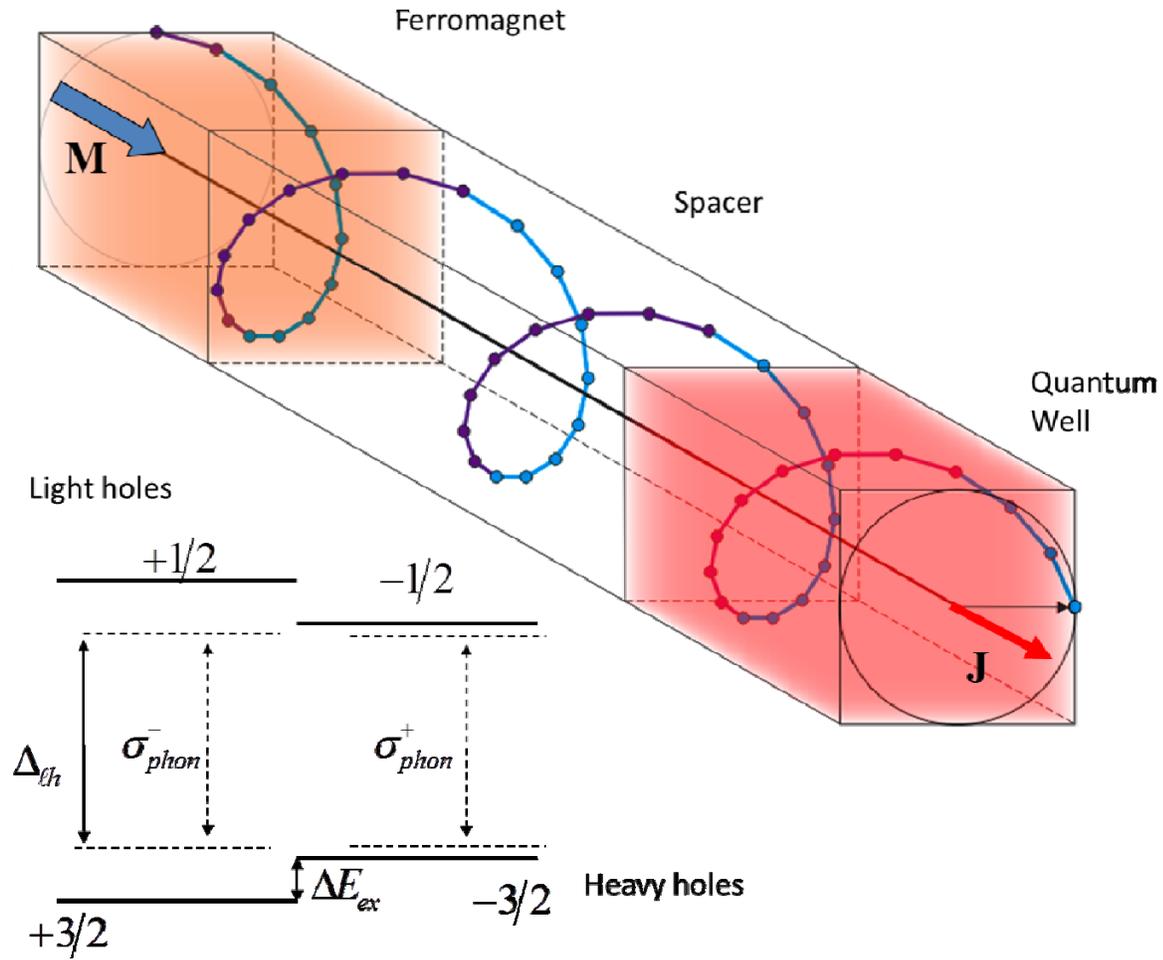

**Figure 5. Illustration of circularly polarized phonon mode in the FM-QW hybrid structure.** Energy diagram: a circularly polarized phonon couples the ground state of heavy hole and the excited state of light hole inducing spin-dependent shift of hole spin levels – "phonon ac Stark effect".

**Supplementary information**

# Long-range p-d exchange interaction in a ferromagnet-semiconductor hybrid structure

V. L. Korenev[1,2], M. Salewski[2], I. A. Akimov[1,2], V. F. Sapega[1,3], L. Langer[2], I. V. Kalitukha[1], J. Debus[2], R. I. Dzhioev[1], D. R. Yakovlev[1,2], D. Müller[2], C. Schröder[4], H. Hövel[4], G. Karczewski[5], M. Wiater[5], T. Wojtowicz[5], Yu. G. Kusrayev[1], and M. Bayer[1,2]

[1]Ioffe Physical-Technical Institute, Russian Academy of Sciences, 194021 St. Petersburg, Russia

[2]Experimentelle Physik 2, Technische Universitat Dortmund, D-44227 Dortmund, Germany

[3]Physical Faculty of St. Petersburg State University, 198504 St. Petersburg, Russia

[4]Experimentelle Physik 1, Technische Universitat Dortmund, D-44227 Dortmund, Germany

[5]Institute of Physics, Polish Academy of Sciences, PL-02668 Warsaw, Poland

## A. Origin of the ferromagnet-induced QW PL circular polarization.

Here we discuss more possibilities of FM-induced circularly polarized emission from the QW. We show that these alternative mechanisms cannot explain the experimental data.

### A1. Effect of magnetic circular dichroism.

We can exclude any potential influence from magnetic circular dichroism (MCD), i.e., the dependence of the absorption coefficient on photon helicity. MCD would induce a PL polarization even if the carriers in the QW would be non-polarized. To check, detection was done at the e-$A^0$ PL transition (1.494 eV) of the GaAs substrate, and the laser was scanned with alternating helicity across the QW resonances in the range from 1.59 to 1.63 eV with a Faraday-field of $B_F = \pm 70$ mT applied. Only negligible changes of the PL intensity <0.3% were found. Therefore MCD is not important and the observed $A \approx 4\%$ polarization results from the ferromagnet-induced spin polarization of two-dimensional electrons and/or holes.

### A2. Role of stray fields in the FM-induced spin polarization.

Optical orientation of electrons occurs under excitation by circularly polarized light. The dependencies of the circular polarization degree $\rho_c^\sigma(B)$ on magnetic field in Voigt geometry (the Hanle effect) and in Faraday geometry (effect of stabilization of the optical orientation, recently called the "inverted" Hanle effect [S1]) are determined by the distribution of magnetic stray fields [S2]. The Hanle effect (Fig. S1) is the depolarization of PL in a magnetic field $B_V$ in Voigt geometry and results from the electron spin precession about $B_V$. The transversal hole g-factor is close to zero so that the hole depolarization is negligible. The half-width of the depolarization curve $B_{1/2}^V \approx 15$ mT in structure 1 for the 7-10 nm range of the spacer thickness. A magnetic field in Faraday geometry increases the circular polarization ("inverted" Hanle effect) with a similar characteristic field $B_{1/2}^F \approx 30$ mT. This behavior points towards the presence of stray magnetic fields in the QW region due to the domain structure of the FM [S2] or interface roughness [S1]. (Note that the fringe fields due to hyperfine interaction with the nuclei are about 0.5 mT in CdTe QW [S3] and can be safely neglected). The effect of the stray fields is also seen in TRPL data, where they lead to a dephasing of the optical orientation of electrons with a characteristic time $\tau_{deph} = T_2^* = (1.0 \pm 0.3)$ ns, see the lower panel of Fig. 3d. For the case of an isotropic distribution of the random fields there is a relation [S4] between $B_{1/2}^V$, $B_{1/2}^F$, and $T_2^*$:

$B_{1/2}^F = B_{1/2}^V = 2\sqrt{3}\hbar/\mu_B g_e T_2^*$. Taking into account $|g_e| = 1.2$ one obtains $B_{1/2}^{F/V} = 22$ mT, which is between the $B_{1/2}^V \approx 15$ mT and $B_{1/2}^F \approx 30$ mT observed experimentally. The difference from $B_{1/2}^V \neq B_{1/2}^F$ indicates an anisotropic distribution of the FM random fields. Small local fields in the ten mT range can affect the electron spin precession kinetics. However, the polarization of electrons (and of holes whose longitudinal g-factor is similar) due to thermal distribution between the spin sublevels at $T = 2$ K cannot exceed 0.5% at this field strength, i.e. is too small to explain experiment.

### A3. Role of diffusion of magnetic atoms into the QW region or of cluster formation

Diffusion of magnetic atoms into the QW region could enhance the effective Lande g-factors of spins, so that $g_{eff} \gg 1$, due to exchange interaction similar to the behavior in diluted magnetic semiconductors. Another possibility for the small saturation field $B_{sat} \sim 20$ mT of $\rho_c^\pi(B_F)$ (see Fig. S2) could be superparamagnetic Co cluster formation. For clarification, the polarization dependencies $\rho_c^\pi(B_F)$ were measured at different temperatures in the range from 2 up to 30 K. Figure S2 shows that all curves coincide when the measured amplitudes are scaled accordingly. Therefore the polarization originates from a ferromagnet, and not from paramagnetic clusters or Co impurities. In the latter cases, the saturation of $\rho_c^\pi(B)$ would depend on the ratio $M_c B/k_B T$ ($k_B$ is the Boltzmann constant, and $M_c$ is the magnetic moment of the cluster) or on $\mu_B g_{eff} B/k_B T$ for Co impurities. With increasing temperature polarization saturation should occur then in larger fields, $B_{sat}(T) \sim k_B T$. However, the saturation of $\rho_c^\pi(B_F)$ takes place at $B_{sat} = 20$ mT over the *entire* temperature range in Fig. S2. This also means that the ferromagnetism of the Co layer is not sensitive to temperature up to 30 K (the origin of the ferromagnetic behavior will be discussed below). The decrease of polarization with increasing temperature (Fig. 4a) is related to the decreasing sensitivity of the "QW detector" to ferromagnetism.

**B. Origin of the ferromagnet responsible for the spin polarization of the QW holes.**

The magneto-optical Kerr effect is often used to measure the magnetic hysteresis loop of thin magnetic films. In our case, the Kerr data show that the magnetization of the Co film is oriented in-plane due to demagnetization with the saturation field $B_F$ being about 1.5 T (see Fig. S3a), which is a typical value for $4\pi M_s$ in Co. The magnetization by an in-plane magnetic field $B_V$ perpendicular and parallel to the spacer gradient direction show magnetic hysteresis loops with a coercivity $B_c = 120$ mT (see Fig. S3b). No changes within the 20 mT accuracy can be seen in Figs. S3a and S3b. These data are in contrast with the proximity effect: The PL polarization degree $\rho_c^\pi$ in perpendicular magnetic field $B_F$ saturates at $B_F = 20$ mT (Fig. 4b of the main text and Fig. S1) and is weakly sensitive to an in-plane magnetic field with $B_V < 100$ mT.

We conclude that the Kerr technique and the PL polarization technique detect ferromagnets with different properties – the easy-plane for the former and the perpendicular anisotropy for the latter method. The FM detected by the Kerr effect can be ascribed to the Co film itself, whereas the other FM could be related to interfacial ferromagnetism with properties differing substantially from the Co film. The interfacial FM layer is expected to be thinner than the Co film thickness, so that it is hard to be seen by the Kerr technique. In contrast, the FM-QW exchange coupling may be sufficient to induce spin polarization of nearby QW charge carriers. Interfacial ferromagnetism was discovered in a Ni/GaAs hybrid [S2] and was discussed later in a number of papers [S5, S6]. At this stage, the origin of the FM responsible for the proximity effect in the Co/(Cd,Mg)Te/CdTe hybrids studied here needs further investigation in future. Currently we can exclude Co-related oxides as origin because the structures 2 and 3 were grown without exposure to air before FM sputtering. Therefore, possible origins are Co-(Cd,Mg)Te intermixing and/or interfacial ferromagnetism. Interestingly, the FM proximity effect decreases with increasing Co thickness. Figure B3 shows the time-integrated amplitude $A=\rho_c^\pi(40\,\text{mT})$ versus the Co thickness $d_{Co}$ for a fixed spacer thickness $d_s = 10\,\text{nm}$ at $T=2\,\text{K}$. The proximity effect decreases not only for small Co thicknesses $d_{Co} < 4\,\text{nm}$ (which is natural) but also for $d_{Co} > 4\,\text{nm}$. The latter may be related to reorientation of the easy axis or change of the interfacial ferromagnetism, when the Co film becomes thick enough.

**C. Hole spin – phonon coupling in the semiconductor valence band**

The interaction of the hole spin **J** with acoustic phonons through the shear strain is described by the Hamiltonian [S7]

$$H_{ph-h} = \frac{2d}{\sqrt{3}}\left([J_z,J_x]_s \varepsilon_{xz} + [J_z,J_y]_s \varepsilon_{yz} + [J_y,J_x]_s \varepsilon_{yx}\right)$$

where $[J_z,J_x]_s = (J_z J_x + J_x J_z)/2$, and the constant $d$ has a typical value of about 10 eV. We assume for simplicity that the Hamiltonian has no dependence on $x$, $y$. Therefore, the components of the strain tensor $\varepsilon_{xz} = \frac{1}{2}\frac{\partial R_x}{\partial z}$, $\varepsilon_{yz} = \frac{1}{2}\frac{\partial R_y}{\partial z}$, $\varepsilon_{xy} = 0$. The $\alpha = x, y$ components of the displacement vector are given by $R_\alpha = \sum_k \sqrt{\frac{\hbar}{2\rho\omega_k}}\left(\xi_\alpha b_k e^{+ikz} + \xi_\alpha^* b_k^+ e^{-ikz}\right)$, where $b_k^+(b_k)$ are the creation (destruction) operators of a phonon with momentum $k$ and frequency $\omega_k$, propagating along the $z$ axis, $\xi_\alpha(\xi_\alpha^*)$ is the complex $\alpha$-component of the polarization vector $\vec{\xi}$, $\rho$ is the mass density of the crystal. The Hamiltonian takes a convenient form after transformation to the ladder operators $J_\pm = J_x \pm iJ_y$ and the circularly polarized basis

$$R_\pm = R_x \pm iR_y = \sum_k \sqrt{\frac{\hbar}{2\rho\omega_k}}\left[(\xi_x \pm i\xi_y)b_k e^{+ikz} + (\xi_x \pm i\xi_y)^* b_k^+ e^{-ikz}\right]$$

$$H_{ph-h} = \frac{d}{2\sqrt{3}} \left( [J_z, J_-]_s \frac{\partial R_+}{\partial z} + [J_z, J_+]_s \frac{\partial R_-}{\partial z} \right) \tag{C1}$$

The first term in Eq. (C1) couples the ground state $|+3/2, N\rangle$ of the heavy hole in the QW in presence of $N$ phonons to the excited state $|+1/2, N-1\rangle$ of light holes in presence of $N$-1 phonons with a probability amplitude $\langle +1/2, N-1|H_{ph-h}|+3/2, N\rangle$ via a $\sigma^-_{phon}$ circularly polarized phonon (coefficients $\xi_x = 1/\sqrt{2}, \xi_y = -i/\sqrt{2}$), in accordance with angular momentum conservation. Similarly, the second term in Eq. (C1) couples the ground state $|-3/2, N\rangle$ of the QW heavy hole in presence of $N$ phonons and the excited light hole state $|-1/2, N-1\rangle$ in presence of $N$-1 phonons with the probability amplitude $\langle -1/2, N-1|H_{ph-h}|-3/2, N\rangle$ via a $\sigma^+_{phon}$ phonon. As the number of phonons with opposite circular polarizations is different especially near the magnon-phonon resonance, the probability amplitudes for the two couplings $+3/2 \leftrightarrow +1/2$ and $-3/2 \leftrightarrow -1/2$ are different. Assuming that near the resonance only the $\sigma^-_{phon}$ phonon mode survives, we obtain a splitting of the heavy-hole states due to the "phonon ac Stark effect"

$$\Delta E_{ex} = \sum_q \frac{\left|\langle +1/2, N_q-1|H_{ph-h}|+3/2, N_q\rangle\right|^2}{\hbar\omega_q - \Delta_{\ell h}} \tag{C2}$$

where the sum is limited to wave vectors $q$, for which the dispersion relation of the transverse acoustic phonons is close to the magnon-phonon resonance, $\hbar\omega_q \approx \hbar\omega_1, \hbar\omega_2$. It follows from Eq.(C2) that the splitting $\Delta E_{ex} = |E_{+3/2} - E_{-3/2}|$ is maximal when the energy of the magnon-phonon resonance $\hbar\omega_1, \hbar\omega_2 < 1\,\text{meV}$ is close to $\Delta_{\ell h}$. For the acceptor the energy splitting $\Delta_{\ell h} \leq 1\,\text{meV}$ is indeed close to the phonon resonance, in contrast to the few ten meV energy splitting between the free hole subbands. Therefore, the spin splitting of the ±3/2 acceptor levels is expected to be larger than that of the QW exciton, in accordance with the stronger polarization for the e-$A^0$ line. Using Eq. (C2) we can estimate the minimal strain component value $\varepsilon$ to produce a splitting $\Delta E_{ex} = 50\,\mu eV$ as estimated in experiment. One obtains $\Delta E_{ex} \sim (d\cdot\varepsilon)^2/\delta$, with a detuning $\delta = \hbar\omega_{1,2} - \Delta_{\ell h}$ from the magnon-phonon resonance. This gives $\varepsilon \sim 10^{-5}$ for $d = 7$ eV and $\delta = 0.1$ meV. Such a small strain value can easily be obtained in semiconductors.

One cannot exclude the possibility of elliptically polarized transversal optical phonons (corresponding to the zero-point motion in the low temperature case). Coupling with circularly polarized optical phonons looks like spin-dependent polaronic shifts and can give a similar splitting value for the free hole and for the hole bound to acceptor. Circularly polarized optical phonons were observed in magnetic field (Ref. 15 of the main text). To illustrate this point we consider for simplicity a crystal with two atoms per unit cell. Uniform strain is characterized by the relative displacement $\mathbf{r} = \mathbf{R}_1 - \mathbf{R}_2$ of the sublattices 1 and 2. Unlike the acoustic phonon case, optical oscillations do not shift the center of mass of the unit cell. Hence the spin-phonon Hamiltonian includes the displacement itself, without derivatives with respect to the coordinates (being typically much smaller)

$$H_{opt-h} = Q_{xyz} r_x [J_z, J_y]_s + Q_{yzx} r_y [J_z, J_x]_s + Q_{zxy} r_z [J_y, J_x]_s \tag{C3}$$

The third rank tensor $\hat{Q}$ has only one independent component $Q = Q_{xyz} = Q_{yzx} = Q_{zxy}$ (dimensionality eV/A) in CdTe-type bulk semiconductors. The order of magnitude of the parameter is $Q \sim e^2/a_0^2 \sim 1$ eV/A ($a_0$ is the linear size of the unit cell). Optical phonon modes propagating along $\mathbf{z} \| \vec{m}$ (the structure growth axis) have angular momentum parallel to *z*. The angular momentum is related to the *x* and *y* components of the displacement vector $\mathbf{r}$. For this reason we omit the last term in Eq. (C3). Similar to (C1) it is convenient to transform the Hamiltonian (C4) to the operators $J_\pm = J_x \pm iJ_y$ and the circularly polarized basis

$r_\pm = r_x \pm ir_y = \sum_k \sqrt{\dfrac{\hbar}{2\rho\omega_0}} \left[ \left(\xi_x^0 \pm i\xi_y^0\right) b_k e^{+ikz} + \left(\xi_x^0 \pm i\xi_y^0\right)^* b_k^+ e^{-ikz} \right]$ where $\vec{\xi}^0$ is the complex polarization vector of the optical phonon, and $\omega_0$ is its frequency. One obtains

$$H_{opt-h} = \frac{Q}{2i}\left(r_+[J_z,J_+]_s - r_-[J_z,J_-]_s\right) \tag{C4}$$

We consider the low-temperature case $\Delta_{\ell h}, \hbar\omega_0 \gg k_B T$, when only +3/2 heavy hole states are populated and there are no optical phonons. The shift of the energy level of the +3/2 hole follows from second-order perturbation theory

$$E_{+3/2} = -\sum_q \frac{\left|\langle +1/2, 1_q | H_{opt-h} | +3/2, 0_q \rangle\right|^2}{\hbar\omega_0^+ + \Delta_{\ell h}} \tag{C5}$$

where the matrix element in Eq. (C5) couples the ground +3/2 state without phonons with the excited +1/2 hole state with one phonon $1_q$ having momentum **q** along **z** and $\sigma^+$ polarization (the second term in (C4) is responsible for this coupling). The upper index in the phonon frequency takes explicitly into account the time-reversal broken symmetry in the FM/SC hybrid, so that the phonon energies $\hbar\omega_0^+ \neq \hbar\omega_0^-$ are different for $\sigma^+$ and $\sigma^-$ polarizations. Using Eq. (C5) we obtain for a small splitting $|\hbar\omega_0^+ - \hbar\omega_0^-| \ll \hbar\omega_0$ of the phonon mode

$$\Delta E_{ex}^{opt} = E_{+3/2} - E_{-3/2} = \beta\left(\hbar\omega_0^+ - \hbar\omega_0^-\right) \tag{C6}$$

where the coefficient $\beta \sim |\langle 1|H_{opt-h}|0\rangle|^2/(\hbar\omega_0)^2 \ll 1$. We do not know the difference $|\hbar\omega_0^+ - \hbar\omega_0^-|$ and the parameter $\beta$ for the studied hybrid. Taking as a reference point $|\hbar\omega_0^+ - \hbar\omega_0^-| = 20$ cm$^{-1}$ [15] and our estimation $\Delta E_{ex} = 50 \mu eV$ one has $\beta \sim 0.01$. Finally we note that the optical phonon energy $\hbar\omega_0 \approx 21$ meV is larger than the $\Delta_{\ell h}$ splitting for both the valence band heavy hole (about 10 meV) and the hole bound to a shallow acceptor (about 1 meV). Hence the splitting $\Delta E_{ex}^{opt}$ will be comparable for both hole states in contrast to the case of interaction with acoustic phonons (C2).

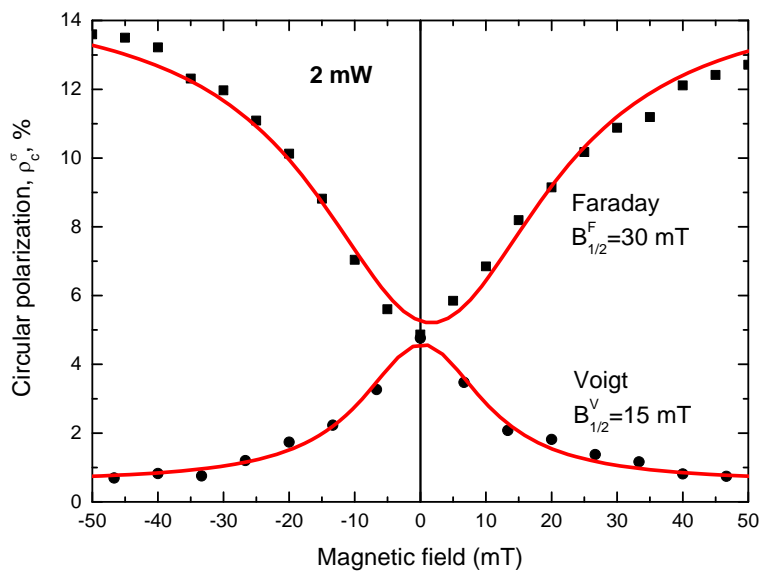

**Figure S1. Magnetic field dependence of circular polarization.** Hanle effect (the circles) in Voigt geometry, fitted with a Lorentzian with $B_{1/2}^{V} \approx 15$ mT. "Inverted" Hanle effect (the squares), also fitted by a Lorentzian with $B_{1/2}^{F} \approx 30$ mT; $T = 10$ K, structure 1.

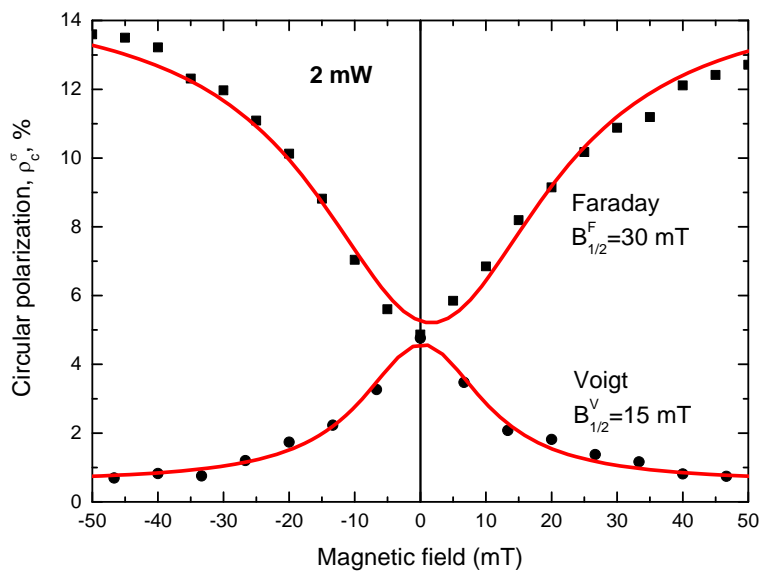

**Figure S1. Magnetic field dependence of circular polarization.** Hanle effect (the circles) in Voigt geometry, fitted with a Lorentzian with $B_{1/2}^{V} \approx 15$ mT. "Inverted" Hanle effect (the squares), also fitted by a Lorentzian with $B_{1/2}^{F} \approx 30$ mT; $T = 10$ K, structure 1.

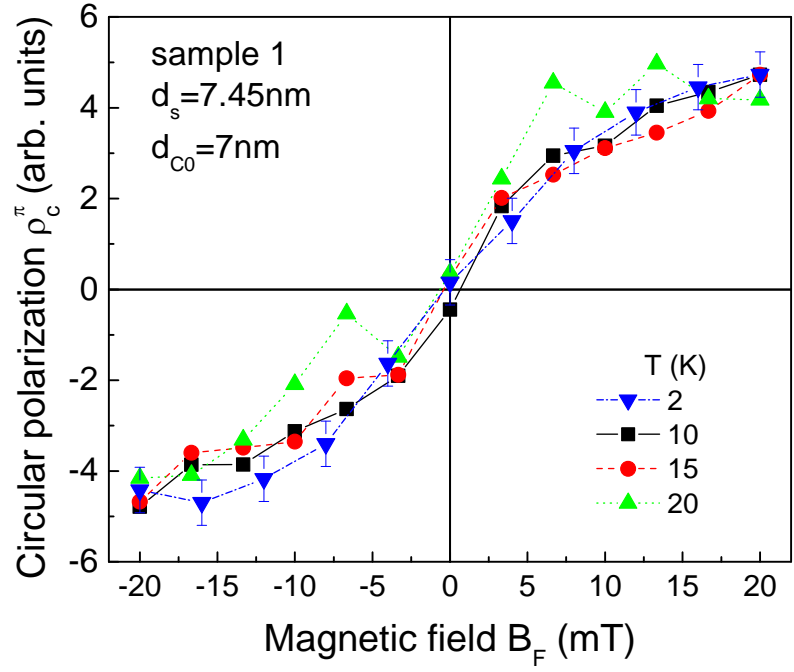

**Figure S2. Effective *p-d* exchange interaction between FM and QW heavy holes.** Dependence $\rho_c^\pi(B_F)$ on magnetic field at different temperatures for a spacer thickness of 7.5 nm; all curves are rescales to a single one by rescaling the vertical scale of the data to the same average circular polarization value.

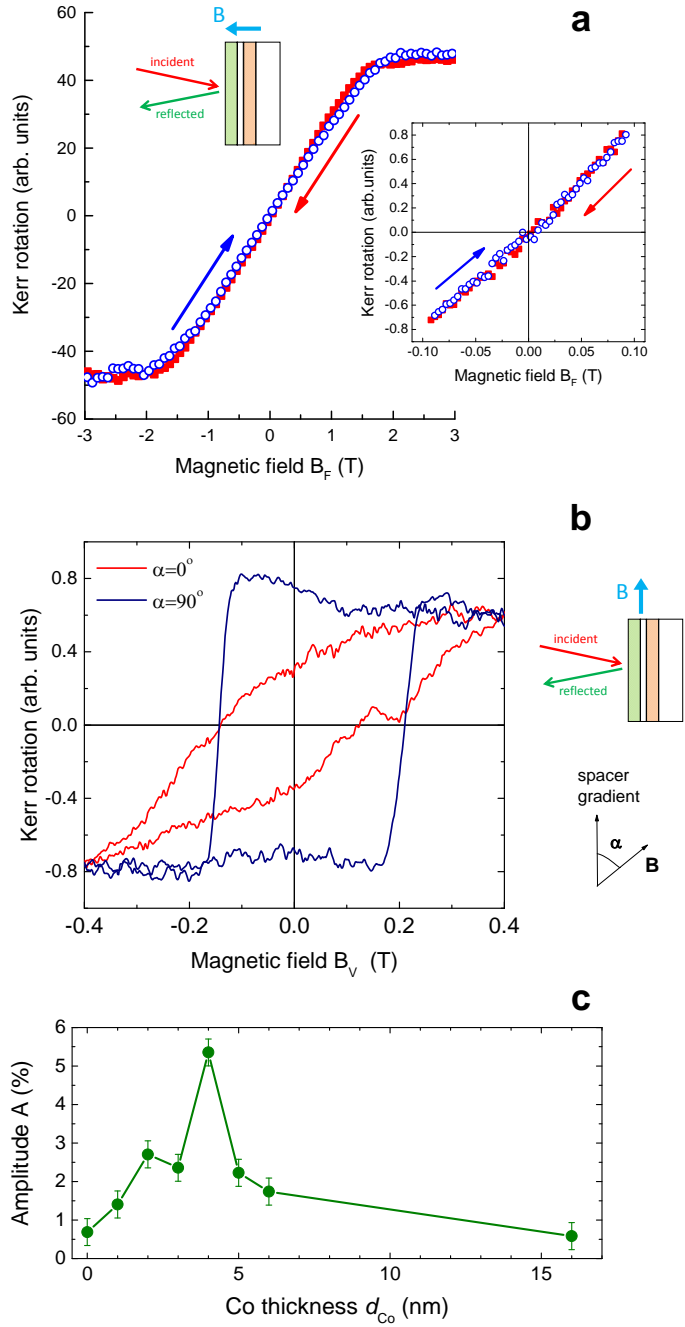

**Figure S3.** (a) The polar Kerr effect reveals the out-of plane magnetization component in Faraday field $B_F$. (b) The longitudinal Kerr effect detects the in plane magnetization component for two different $B_V$ field orientations, parallel and perpendicular to the spacer thickness gradient. (c) Time-integrated amplitude $A=\rho_c^\pi(B_F=40\,\text{mT})$ versus Co thickness $d_{Co}$ for $d_s=10$ nm and $T=2$ K.

References SOM